\documentclass[traditabstract]{aa}
\usepackage{graphicx}
\usepackage{txfonts}

\def\tex {\ifmmode{{T}_{\rm ex}}\else{$T_{\rm ex}$}\fi}
\def\tmb {\ifmmode{{T}_{\rm mb}}\else{$T_{\rm mb}$}\fi}
\def\ci     {\ifmmode{{\rm C}{\rm \small I}}\else{C\ts {\scriptsize I}}\fi}
\def\hi     {\ifmmode{{\rm H}{\rm \small I}}\else{H\ts {\scriptsize I}}\fi}
\def\hh     {\ifmmode{{\rm H}_2}\else{H$_2$}\fi}

\def\ts     {\thinspace}
\def\kms    {\ifmmode{{\rm \ts km\ts s}^{-1}}\else{\ts km\ts s$^{-1}$}\fi}
\def\msol   {\ifmmode{{\rm M}_{\odot}}\else{M$_{\odot}$}\fi}
\def\lsol   {\ifmmode{{\rm L}_{\odot}}\else{L$_{\odot}$}\fi}
\def\zsol   {\ifmmode{{\rm Z}_{\odot}}\else{Z$_{\odot}$}\fi}
\def\etal   {{\rm et\ts al.}~}

\begin{document}

\title{ALMA reveals the feeding of the Seyfert 1 nucleus in NGC 1566
\thanks{Based on observations carried out with  ALMA in cycle 0.
}}

\author{F. Combes \inst{1}
\and
S. Garc\'{\i}a-Burillo \inst{2}
\and
V. Casasola \inst{3}
\and
L. K. Hunt \inst{4}
\and
M. Krips \inst{5}
\and
A. J. Baker \inst{6}
\and
F. Boone \inst{7}
\and
A. Eckart \inst{8}
\and
I. Marquez \inst{9}
\and
R. Neri \inst{5}
\and
E. Schinnerer \inst{10}
\and
L. J. Tacconi \inst{11}
           }
\offprints{F. Combes}
\institute{Observatoire de Paris, LERMA (CNRS:UMR8112), 61 Av. de l'Observatoire, F-75014, Paris, France
\email{francoise.combes@obspm.fr}
 \and
Observatorio Astron\'omico Nacional (OAN)-Observatorio de Madrid,
Alfonso XII, 3, 28014-Madrid, Spain
 \and
INAF -- Istituto di Radioastronomia \& Italian ALMA Regional Centre, via Gobetti 101, 40129, Bologna, Italy
 \and
INAF - Osservatorio Astrofisico di Arcetri, Largo E. Fermi, 5, 50125, Firenze, Italy
 \and
IRAM, 300 rue de la Piscine, Domaine Universitaire, F-38406 Saint Martin d'H\`eres, France
 \and
Dep. of Physics \& Astronomy, Rutgers, the State University of New Jersey, 136 Frelinghuysen road, Piscataway, NJ 08854, USA
 \and
CNRS, IRAP, 9 Av. colonel Roche, BP 44346, 31028, Toulouse Cedex 4, France
 \and
I. Physikalisches Institut, Universit\"at zu K\"oln, Z\"ulpicher Str. 77, 50937, K\"oln, Germany
 \and
Instituto de Astrof{\'{\i}}sica de Andaluc{\'{\i}}a (CSIC), Apdo 3004, 18080 Granada, Spain
 \and
Max-Planck-Institut f\"ur Astronomie (MPIA), K\"onigstuhl 17, 69117 Heidelberg, Germany
 \and
Max-Planck-Institut f\"ur extraterrestrische Physik, Giessenbachstr. 1, Garching bei M\"unchen, Germany
              }

   \date{Received  2014/ Accepted  2014}

   \titlerunning{CO in NGC 1566}
   \authorrunning{F. Combes et al.}

   \abstract{We report ALMA observations of CO(3-2) emission in the Seyfert 1 galaxy NGC~1566,
at a spatial resolution of 25~pc. Our aim is to investigate the morphology and
dynamics of the gas inside the central kpc, and to probe nuclear fueling and feedback phenomena.
NGC~1566 has a nuclear bar of 1.7~kpc radius and a conspicuous grand design spiral 
starting from this radius. The ALMA field of view, of diameter 0.9~kpc, lies well inside the nuclear bar
and reveals a molecular trailing spiral structure from 50 to 300~pc in size, which is contributing to fuel the nucleus,
according to its negative gravity torques.
The spiral starts with a large pitch angle from the center and then winds up 
in a pseudo-ring at the inner Lindblad resonance (ILR) of the nuclear bar.
 This is the first time that a trailing spiral structure is clearly seen driving the gas inwards inside
the ILR ring of the nuclear bar. This phenomenon shows that the massive central black hole has a significant
dynamical influence on the gas, triggering its fueling. 
 The gaseous spiral is well correlated with the dusty spiral seen through extinction in HST images, and also
with a spiral feature emitting 0.87mm continuum. This 
continuum emission must come essentially from cold dust heated by the interstellar radiation field. 
The HCN(4-3) and HCO$^+$(4-3) lines were simultaneously mapped and detected in the nuclear spiral.
The HCO$^+$(4-3) line is 3 times stronger than the HCN(4-3), as expected when star formation excitation dominates
over active galactic nucleus (AGN) heating. The CO(3-2)/HCO$^+$(4-3) integrated intensity ratio is $\sim$ 100.
 The molecular gas is in remarkably regular rotation, with only slight non-circular motions at the periphery of the nuclear spiral arms. 
These perturbations are quite small, and no outflow nor AGN feedback is detected.
\keywords{Galaxies: active --- Galaxies: Individual: NGC 1566 --- Galaxies: ISM --- Galaxies: kinemativs and dynamics
 --- Galaxies: nuclei --- Galaxies: spiral}
}
\maketitle


\section{Introduction}

An astrophysical problem that has seen much progress in recent years is understanding how active galactic nuclei (AGN) 
are fueled in galaxies, and how the energy generated by the active nucleus can in turn regulate gas accretion 
(e.g. Croton \etal 2006; Sijacki \etal 2007). This has important implications for the co-evolution of galaxies and black holes, 
which is observed through the now well established M-$\sigma$ relation (e.g., G\"ultekin \etal 2009).  Driving sufficient gas towards 
the center to sustain nuclear activity requires the removal of angular momentum from the gas, and therefore the creation of large 
non-axisymmetries. The invoked dynamical mechanisms depend on the scale in question: at 10~kpc scales, torques are produced by 
galaxy interactions and mergers (e.g., Hopkins \etal  2006; di Matteo \etal 2008); at kpc scales, bar instabilities, either internally driven 
by secular evolution or triggered by 
companions, can first feed a central starburst and then fuel the massive black hole (Garc\'{\i}a-Burillo \etal 2005). 
 At 300~pc scales, 
the ``bars within bars'' scenario (e.g., Shlosman \etal 1989), together with m=1 instabilities, takes over as a dynamical mechanism;
 however, our IRAM-PdBI observational program NUclei of GAlaxies (NUGA) has revealed smoking-gun evidence of AGN fueling 
only in one third of the galaxies observed (Garc\'{\i}a-Burillo \& Combes 2012). This might be due to nuclear star formation 
or AGN feedback driving gas outflows that stop the fueling; galaxies will then undergo successive periods of fueling and starvation, 
and might be found in a feeding phase at 300~pc scales only one third of the time.
  Other mechanisms could contribute to the fueling, especially towards the center of galaxies: viscous torques, 
from dense gas in regions of large shear, or dynamical friction of massive clouds against the
old bulge stars (e.g. Combes 2002, Jogee 2006). At the radii of interest, though, these are slower
than gravity torques, when present. 

To go beyond these scales and explore the molecular ISM 30~pc to 300~pc scales in local AGN host galaxies, we have 
undertaken an observing campaign with ALMA, the only instrument able to probe the molecular emission with enough spatial 
resolution and sensitivity. Observations and simulations are advancing together, since it is only now that we can simulate coherently
 these scales with successive zoom-in re-simulations (e.g., Hopkins \& Quataert 2010; Renaud \etal 2013).  
 At 30~pc scales, 
simulations suggest fueling involves a cascade of dynamical instabilities (m=2, m=1), and the formation of a thick gas disk
similar to a torus, subject to 
bending and warping instabilities (Hopkins \etal 2012). Since the gas concentration becomes large, the gas is highly unstable, 
clumpy and turbulent. Dynamical friction can then drive these massive clumps to the nucleus. Simulations reveal episodic feeding 
with 10Myr time-scales (10~pc) or even lower (0.1pc). These fueling episodes are then quenched by either star formation winds 
or AGN outflows.

Recently, high velocity outflows have been discovered in the molecular gas of nearby AGN (Feruglio \etal 2010; Sturm \etal 2011; 
Alatalo \etal 2011; Dasyra \& Combes 2012). Outflows have been traced for a long time in ionized or atomic gas (Rupke \etal 2005; 
Riffel \& Storchi-Bergmann 2011). With our ALMA cycle 0 data,  we have been able to confirm that molecular outflows 
are also seen in low-luminosity AGN. The outflow revealed in the Seyfert 2 NGC~1433 is the least massive molecular outflow ever seen 
in galaxy nuclei (Combes \etal 2013).
In the prototypical Seyfert 2 NGC~1068, a clear molecular ouflow has also been detected, entrained by the 
AGN radio jets (Krips \etal 2011; Garc\'{\i}a-Burillo \etal 2010, 2014 in prep.).

These high angular resolution results have shown that fueling and feedback phases can occur simultaneously. Gas is inflowing 
in the plane, with some also outflowing along the minor axis at some angle with the plane. In NGC 1433, the CO map reveals a 
nuclear gaseous spiral structure inside the 460~pc-radius nuclear ring encircling the nuclear stellar bar. The nuclear spiral winds up 
in a pseudo-ring at $\sim$200 pc radius, which might correspond to the inner Lindblad resonance (ILR). 
In spite of the presence of gas in rings at both outer and inner ILRs, 
some gas is finding its way to the center. In both of these Seyfert 2 galaxies, dust continuum emission is detected in a small nuclear disk, 
which might be interpreted as the molecular torus
(Combes \etal 2013). 
In NGC 1068, the AGN is clearly off-center with respect to the torus, implying an m=1 perturbation
(Garc\'{\i}a-Burillo \etal 2014 in prep.).

In this paper, we present ALMA cycle 0 observations in the CO(3-2) line of the Seyfert 1 NGC~1566, with a spatial
resolution of $\sim$ 25~pc.
Its proximity (10~Mpc) and small inclination of 35$^\circ$ make NGC~1566 an ideal target to test and 
refine scenarios of AGN feeding and feedback, and discover new phenomena controlling
gas structures and dynamics within 100 pc, where the dynamical time-scale is smaller than 3 Myr
(for V$_{rot} \sim$ 200 km/s).
In the next subsection, all relevant characteristics of NGC~1566 are described. Observations are detailed in 
Section \ref{obs} and results in Section \ref{res}. The interpretation in term of torques is discussed in
Section \ref{torq}, and conclusions are drawn in Section \ref{disc}.

\begin{figure}[ht]
\centerline{
\includegraphics[angle=-90,width=8cm]{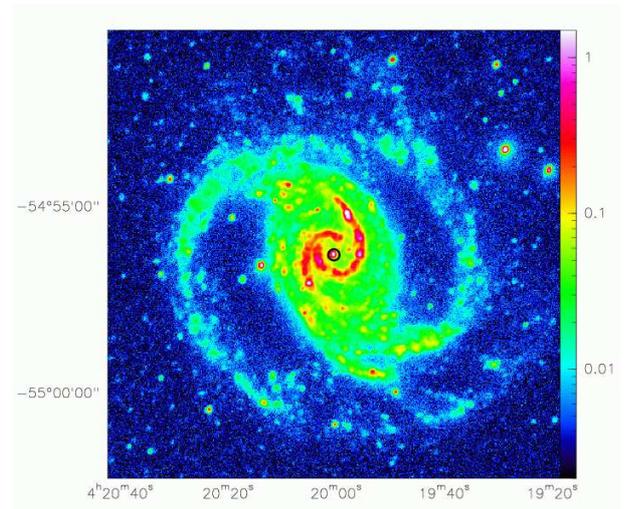}}
\caption{ {\it GALEX}-NUV image of NGC~1566, showing the two sets of spiral arm structures
at large scales. The ALMA field of view (9\arcsec\ radius) is shown as the black circle in the center.
The image is a square of 12\arcmin\ size. The color scale is in log.} 
\label{fig:galex}
\end{figure}

\begin{figure*}[ht]
\centerline{
\includegraphics[angle=-90,width=15cm]{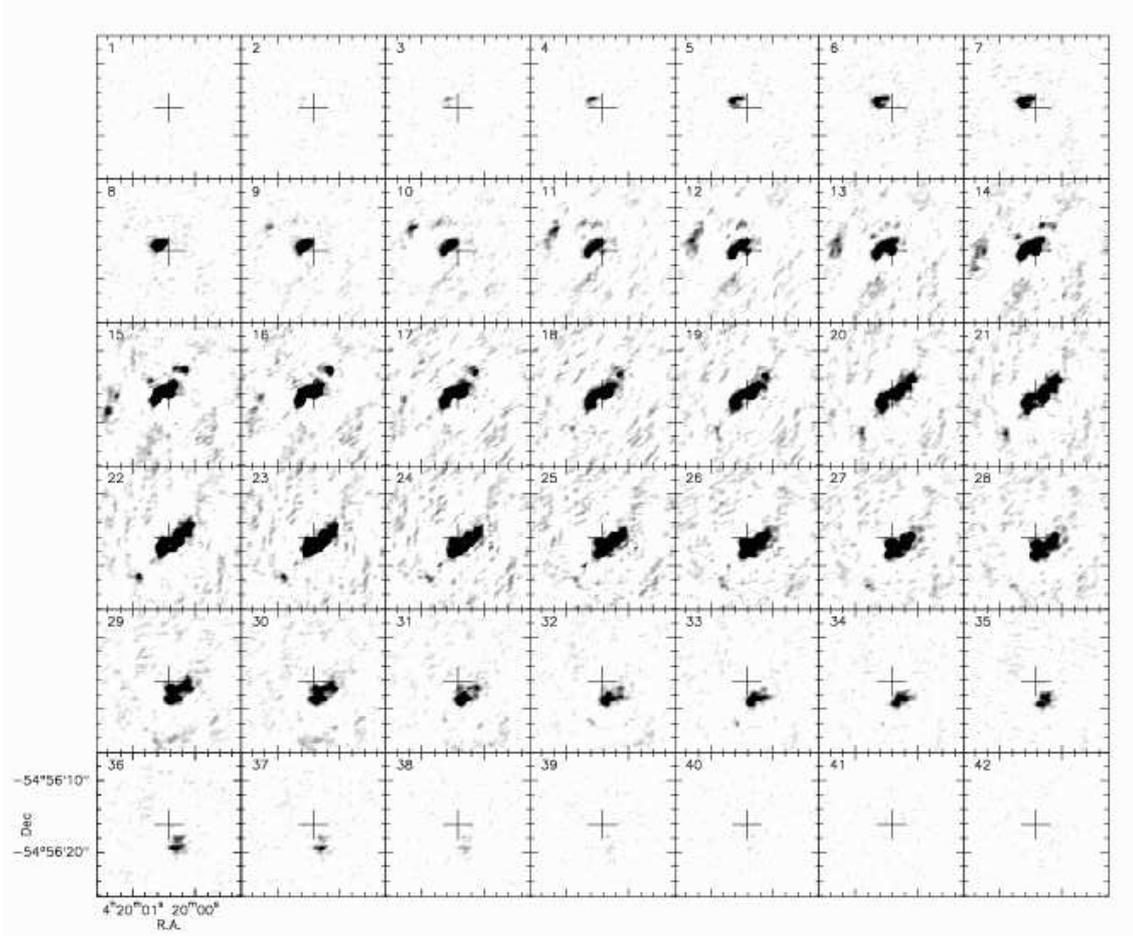}}
\caption{ Channel maps of CO(3-2) emission in the center of NGC~1566.
Each of the 42 panels is 20\arcsec\, in size. Channels are separated
by 10.15~km/s. They are plotted from V$_{hel}$=1302 (top left) to 1718.15~km/s (bottom right).
The synthesized beam is 0.64\arcsec\ x 0.43\arcsec\ (PA=123$^\circ$). The phase center of 
the observations (marked by a 4\arcsec\ cross)
is given in Table \ref{tab:basic}. The grey scale is linear, between 1 and 
30 mJy/beam. The central channels are saturated to better show the extent of the emission.
}
\label{fig:chann}
\end{figure*}

\begin{figure}[ht]
\centerline{
\includegraphics[angle=-0,width=8cm]{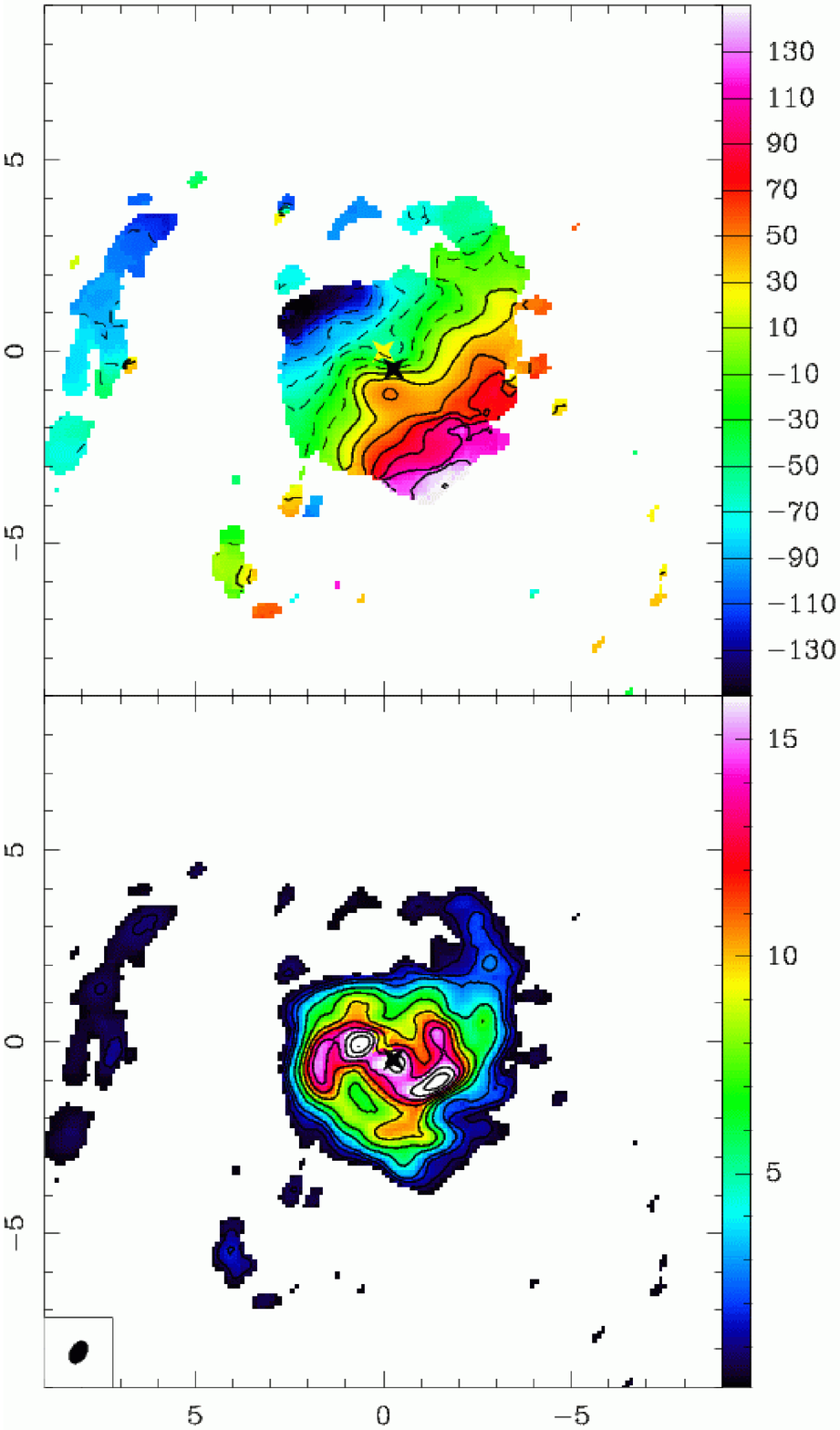}}
\caption{ Velocity field (top) and integrated intensity (bottom) of the CO(3-2) emission in 
the center of NGC~1566. Coordinates are in arcsec relative to the phase center of
Table \ref{tab:basic}. The latter is indicated as a yellow cross, and the new adopted 
center as a black cross (see Sec. \ref{cont}). 
The color palettes are in km/s relative to V$_{hel}$=1504~km/s (top) and 
in Jy/beam$\times$MHz (or 0.87 Jy/beam$\times$km/s) at the bottom. The beam size of 
0.64\arcsec\ x 0.43\arcsec\ is indicated
at the bottom left.}
\label{fig:velo}
\end{figure}

\subsection{NGC 1566}
\label{sample}
NGC~1566, a nearly face-on spiral galaxy, is the brightest member of the Dorado group (Ag\"uero \etal 2004;
Kilborn et al 2005). It has an intermediate strength bar type (SAB), and two strongly contrasted spiral arms,
emanating from the bar and winding up in an outer pseudo ring. The remarkable grand-design spiral
has inspired special studies of its logarithmic arm structure (Elmegreen \& Elmegreen 1990; Korchagin \etal 2000).
There are two sets of spiral arms (see Fig. \ref{fig:galex}): the first set starts at the end of the nuclear
bar of about 35\arcsec=1.7~kpc in radius, better seen in the near infrared
(Hackwell \& Schweizer 1983). These arms then wind up in a ring, with kinks and a circle of star
formation, identified as the corotation of the spiral pattern (Elmegreen \& Elmegreen 1990). A second 
weaker set of spiral arms then winds up in an outer pseudo ring, identified as the outer Lindblad resonance.
The inner Lindblad resonance of the spiral might correspond to the end of the nuclear bar,
and therefore the corotation of the bar. Inside this ILR, there is a deficiency of HII regions
(Comte \& Duquennoy 1982). 
 
NGC~1566 is a low-luminosity AGN, classified as Seyfert, although its precise position between
Seyfert 1 and 2 varies in the literature. More specifically, its nuclear spectrum varies
from Seyfert 1 in its most active phases to Seyfert 2 at minimum activity (Alloin \etal 1985). 
It has many characteristics of Seyfert 1 galaxies/AGN:
a bright central point-like source (Malkan \etal 1998), relatively broad [FeII] line emission in the Broad Line Rregion (Reunanen \etal 2002), 
characteristic variability on scales of months (Alloin \etal 1986),
and an X-ray point-source nuclear luminosity between 2-10~keV of 7 $\times$ 10$^{40}$ erg/s (Levenson \etal 2009).
The nucleus is known to be
variable from X-rays to IR bands (Alloin \etal 1986; Baribaud \etal 1992; Glass 2004). These variations have
been interpreted as instabilities in a thin accretion disk (Abramowicz \etal 1986). 
The mass of the NGC~1566 black hole has been estimated to be 
8.3 $\times$ 10$^6$ \msol\ by the M-$\sigma$ relation and the measured stellar velocity dispersion 
at the center (Woo \& Urry 2002). From its bolometric luminosity, its Eddington ratio
can be estimated as 0.096 (cf Table \ref{tab:basic}). HST [OIII] images show
a one-sided ionization cone towards the south-east (Schmitt \& Kinney 1996). VISIR images
(Reunanen \etal 2010) reveal a dominant central unresolved source with size FWHM $<$ 15 pc at 11.8 $\mu$m. Both
large-scale and nuclear radio emission is seen in NGC~1566 (Harnett 1984, 1987). 
The atomic gas (HI) dynamics have been investigated with the ATCA interferometer by Walsh (2004), 
with a spatial resolution
of $\sim$ 30\arcsec. There is a 30\arcsec x50\arcsec\  HI hole towards the center, where the gas
becomes mostly molecular.  
The CO(1-0) and (2-1) lines have been mapped by Bajaja \etal (1995) with the SEST telescope, 
and are detected out to 60\arcsec radius.
Excited molecular hydrogen has been detected through its ro-vibrational lines at 2$\mu$m (Reunanen \etal 2002).
The  H$_2$ excitation is not by UV fluorescence but through thermal UV heating.
 The contribution from star formation is less than 10\%, and the
dominant H$_2$ excitation mechanism is collisional via shocks. The derived H$_2$ masses observed out to 2\arcsec\  
from the center are 90\msol\ along the ionised cone and 110\msol\ perpendicular to it.

The distance of NGC~1566 is still subject to large uncertainty.
The measured heliocentric velocity
of 1504 km/s, correcting for the solar motion inside the Local Group and accounting for the Virgocentric flow,
yields a distance of 17.8 Mpc (for a Hubble constant of 70 km/s/Mpc). However, 
when the Tully-Fisher relation is used as distance indicator,
the distance is found to be 6.5 Mpc (e.g. Hyperleda). 
For the higher distance, the gas to dynamical mass ratio is unrealistically high, favoring lower values.
In the present work, we adopt the median value given by NED, i.e. D=10 Mpc, for which 1\arcsec\ = 48\,pc.

%
\begin{center}
\begin{table}
      \caption[]{Basic data for the galaxy NGC 1566} 
\label{tab:basic}
\begin{tabular}{lll}
\hline
Parameter  & Value$^{\mathrm{b}}$ & Reference$^{\mathrm{c}}$ \\
\hline
$\alpha_{\rm J2000}$$^{\mathrm{a}}$ & 04$^h$20$^m$00.42$^s$ & (1) \\  
$\delta_{\rm J2000}$$^{\mathrm{a}}$ &-54$^{\circ}$56$^{\prime}$16.1\arcsec\ & (1) \\
$V_{\rm hel}$ & 1504 km\,s$^{-1}$ & (1) \\
RC3 Type & SAB(s)bc & (1) \\
Nuclear Activity & Seyfert 1 & (2) \\
Inclination & 35\fdg0 & (3) \\
Position Angle & 44$^{\circ}$ $\pm$ 1$^{\circ}$ & (3) \\
Distance & 10\,Mpc (1\arcsec\ = 48\,pc) & (1) \\
L$_{B}$        & $8.8 \times 10^{9}$\,L$_{\odot}$ & (4) \\
M$_{\rm H\,I}$ & $3.5 \times 10^{9}$\,M$_{\odot}$ & (5) \\
M$_{\rm H_{2}}$& $1.3 \times 10^{9}$\,M$_{\odot}$ & (6) \\
M$_{\rm dust}$(60 and 100\,$\mu$m)& $3.7 \times 10^{6}$\,M$_{\odot}$ & (7) \\
L$_{\rm FIR}$  & $6.5 \times 10^{9}$\,L$_{\odot}$ & (7) \\
M$_{\rm BH}$  & $8.3 \times 10^{6}$\,M$_{\odot}$ & (2) \\
L$_{\rm bol}$  & $1.0 \times 10^{44}$\,erg/s & (2) \\
$\alpha_{\rm J2000}$$^{\mathrm{d}}$ & 04$^h$20$^m$00.39$^s$ & New center \\
$\delta_{\rm J2000}$$^{\mathrm{d}}$ &-54$^{\circ}$56$^{\prime}$16.6\arcsec\ & New center \\
\hline
\end{tabular}
\begin{list}{}{}
\item[$^{\mathrm{a}}$] ($\alpha_{\rm J2000}$, $\delta_{\rm J2000}$) is the
phase tracking center of our $^{12}$CO interferometric observations.
\item[$^{\mathrm{b}}$]
Luminosity and mass values extracted from the literature
have been scaled to the distance of $D$ = 10\,Mpc.
\item[$^{\mathrm{c}}$] References: 
(1) NASA/IPAC Extragalactic Database (NED, http://nedwww.ipac.caltech.edu/);
(2) Woo \& Urry (2002);
(3) Ag\"uero \etal (2004);
(4) HyperLeda;
(5) Reif \etal\ (1982);
(6) Bajaja \etal\ (1995);
(7) IRAS Catalog.
\item[$^{\mathrm{d}}$]
New adopted center, coinciding with the continuum peak (see Sec. \ref{cont}).
\end{list}
\end{table}
\end{center}

\section{Observations}
\label{obs}

The observations were carried out on October 22, 2012 with the ALMA array in cycle 0,
using 23 antennae, under the project ID 2011.0.00208.S (PI: Combes).
NGC~1566 was observed with Band 7 simultaneously in CO(3-2), HCO$^+$(4-3), 
HCN(4-3), and continuum. The corresponding sky frequencies were
344.09~GHz, 354.98~GHz, 352.76~GHz, and 342.80~GHz, respectively. The observations
were done in 2 blocks, with a total duration of 2 hours.
For each period, NGC~1566 was observed for 28~minutes, and the median
system temperatures were T$_{sys}$ = 149 and 184~K.

The observations were centered on the nucleus, with a single-pointing field of
view (FoV) equal to 18\arcsec; the extended configuration provided
in Band 7 a beam of 0.64\arcsec x 0.43\arcsec, with a PA of 123$^\circ$. 
The galaxy was observed
in dual polarisation mode with 1.875~GHz total bandwidth per baseband, and
velocity resolution of 0.49~MHz $\sim$0.4 km/s. The spectra were then smoothed
to 11.7~MHz (10.15~km/s) to build channel maps.

The correlator configuration was constrained by the allowed
spacing of the intermediate frequencies, and therefore the lines could not be centered
in their bandwidths. Although the bandwidth provided almost 1600 km/s each,
the lines were centered at 220-250~km/s from the edge for HCN and CO, and 440~km/s for HCO$^+$.
Since the total line-width at zero intensity is less than 400~km/s,
this constraint has no impact on the data.
 We stress that we cannot have missed any outflow similar to what have been already
detected in the literature (for instance that in NGC 1266, Alatalo \etal 2011): these flows are 
roughly symmetric in velocity, and should appear complete on the side of full velocity coverage, 
and also as a truncated emission on the other side of the band, which we do not see. 

The total integration time provided an rms of 0.05 mJy/beam in the continuum,
and $\sim$1.3 mJy/beam in the line cubes at 10~km/s resolution. 
Flux calibration was done on Neptune, after subtracting the
line contribution from its spectrum;  band-pass calibration used 
J1924-292, and the phase calibrations the nearby radio source J0455-462. 

The maps were made with Briggs weighting and a robustness parameter of 0.5, i.e. 
a trade-off between uniform and natural weighting.
The data were cleaned using a mask made from the integrated CO(3-2) map.
The continuum was subtracted from all line maps.
The final cubes are 256x256 pixels of 0.11\arcsec\ per pixel in the plane
of the sky, and 50 channels of $\sim$10~km/s width.
The data were calibrated with the CASA software (v4.1; McMullin \etal 2007), and the 
cleaning and imaging
were then finalised with the GILDAS software (Guilloteau \& Lucas 2000).

The maps and images shown in the figures of the present paper are not corrected for primary beam attenuation.
However, the integrated spectra include primary beam correction,
and the total masses are computed with the correction applied. 
Almost no CO(3-2) emission was detected outside the full width at half-power (FWHP) primary beam. 
Due to  missing short spacings, extended emission
was filtered out at scales larger than $\sim$3\arcsec\,  in each 
channel map. The elongated features detected at each velocity
are, however, quite narrow spatially, thinner than 
2\arcsec, and the loss of flux may not be severe at any velocity,
as can be seen in Fig. \ref{fig:chann}.

\begin{table}
      \caption[]{Line fluxes, after primary beam correction. }
\label{tab:line}
\begin{center}
\begin{tabular}{lcccc}
\hline
\scriptsize{Line}& Area &  V & $\Delta$V$^{(1)}$ & Flux$^{(2)}$ \\
     &  Jy km/s &   km/s &   km/s  & Jy \\
\hline
CO(3-2)&  596$\pm$ 2  & 1516$\pm$0.3 &155$\pm$ 1 &3.6 \\
 C1       &   21$\pm$ 1  & 1368$\pm$1   & 42$\pm$ 1 &0.5 \\
 C2       &  197$\pm$ 5 & 1455$\pm$1   & 77$\pm$ 1 &2.4 \\
 C3       &  368$\pm$ 5  & 1545$\pm$1   & 91$\pm$ 1 &3.8 \\
HCO$^+$(4-3) &   5.7$\pm$ 0.6  & 1509$\pm$7 &135$\pm$ 14 &0.04 \\
HCN(4-3)    &   1.8$\pm$ 0.4  & 1540$\pm$7 &57$\pm$ 14 &0.03 \\
\hline
\end{tabular}
\end{center}
\begin{list}{}{}
\item[]   Results of the Gaussian fits: first line for CO(3-2) assumes only one component;
following lines assume 3 velocity components (C1, C2 and C3), shown in Fig. \ref{fig:spectot}.
\item[] $^{(1)}$ Full width at half maximum (FWHM) 
\item[] $^{(2)}$ Peak flux 
\end{list}
\end{table}

\begin{figure}[h!]
\centerline{
\includegraphics[angle=-90,width=8cm]{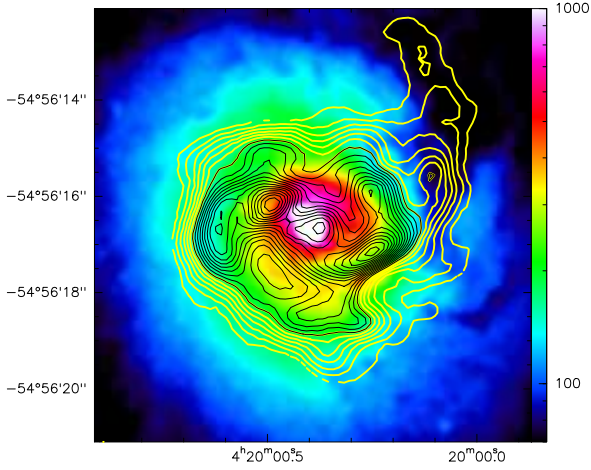}
}
\caption{Overlay of CO(3-2) contours on the  (F606W) HST image.
The colour scale is linear, in arbitrary units. Notice the perfect coincidence between
the emergent spiral arm in CO(3-2) and the dust extinction in the NW.
CO contours are linear from 0.04 to 17.44 in steps of 0.87 Jy/beam.km/s.
The first contours of the CO(3-2) emission are drawn in yellow to
be visible against the dark dust lane.
}
\label{fig:CO-on-V}
\end{figure}

\begin{figure}[h!]
\centerline{
\includegraphics[angle=-90,width=8cm]{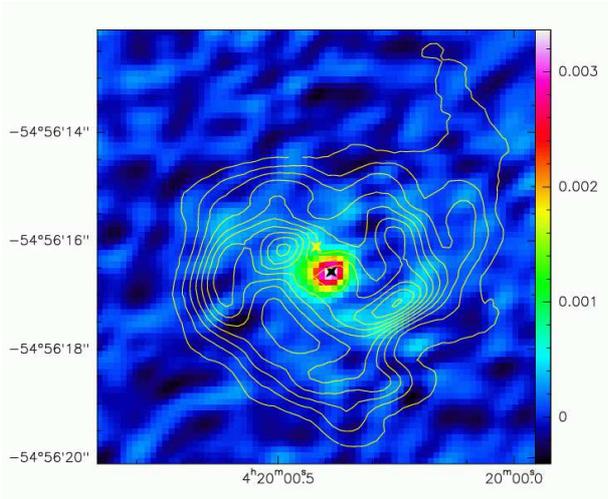}
}
\caption{Overlay of CO(3-2) contours (as in Fig.  \ref{fig:CO-on-V})
on the 0.87mm continuum image. The field of view is
8\arcsec\ in diameter. The yellow cross shows the phase center, while the black cross
defines the new center adopted in Table \ref{tab:basic}. The colour scale is in Jy/beam.
}
\label{fig:cont}
\end{figure}

\begin{figure}[h!]
\centerline{
\includegraphics[angle=0,width=8cm]{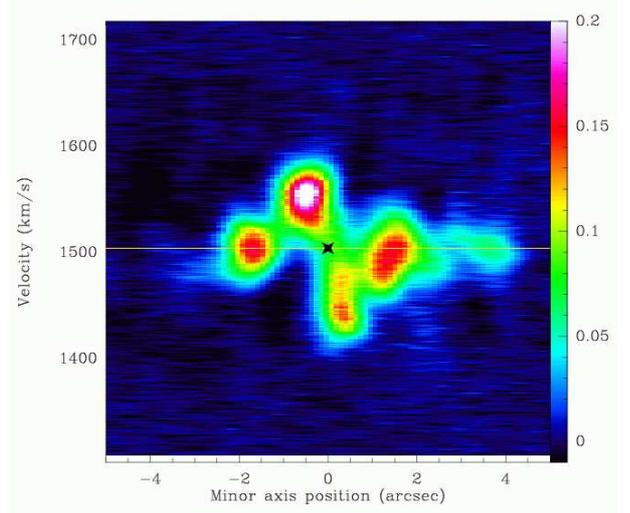}
}
\caption{CO(3-2) position-velocity diagram along the minor axis at PA = 134$^\circ$
(SE is at left, and is the far side). Note the small deviations from a constant radial velocity, which 
come from spiral streaming motions. The black cross indicates the adopted center.
}
\label{fig:pv-minor}
\end{figure}

\begin{figure}[h!]
\centerline{
\includegraphics[angle=0,width=8cm]{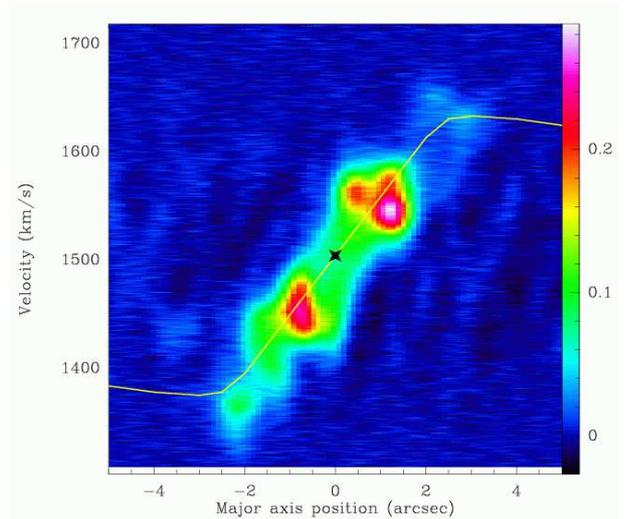}
}
\caption{CO(3-2) position-velocity diagram along the major axis at PA = 44$^\circ$
(East is at left).
The velocity profile is quite regular, and matches with the expected rotational velocity gradient 
(yellow line), as modelled in Fig. \ref{fig:vrot}. The black cross indicates the adopted center.
}
\label{fig:pv}
\end{figure}

\section{Results}
\label{res}

Figure \ref{fig:chann} displays 42 of the CO(3-2) channel maps, with a velocity range
of  416 km/s and a velocity resolution of 10.15 km/s. The velocity field is rather
regular, showing the expected spider diagram, 
although slightly perturbed by spiral structure in the SE and NW directions (see also Fig. \ref{fig:velo}).

\subsection{Molecular gas distribution and morphology}
\label{morpho}

To measure fluxes we used a clipped cube
where all pixel values $<$2$\sigma$ (2.6 mJy/beam) were set to zero. 
The mean intensity is plotted in the lower panel of Fig. \ref{fig:velo}.
Since the galaxy is more extended than the primary beam of 18\arcsec, it is difficult 
to quantify the missing flux. We compare with the central spectrum
obtained with a single dish in Sect. \ref{CO-mass}, and conclude that we might miss
roughly half of the total. These single dish observations
were obtained  with the SEST in CO(2-1) with a 22\arcsec\, beam.
The angular size is comparable to our FoV, but assumptions have to be 
made about the CO line ratios.

The CO emission is concentrated in a nuclear disk of radius 3\arcsec or 150~pc,
where it shows a contrasted 2-arm spiral structure. 
The molecular gas does not seem to follow the nuclear bar. 
 This spiral structure at a $\sim$ 100~pc scale appears independent from the large-scale grand design 
2-arm structure, starting at its pseudo-nuclear ring of $\sim$ 35\arcsec=1.7 kpc in radius
(Elmegreen \& Elmegreen 1990).
We superposed the CO map onto HST images in the U, B, V and I filters (F275W, F450W, F606W, F814W).
All show a remarkable similarity in morphology, with the 
dust extinction features progressively less contrasted, as displayed in Figure \ref{fig:CO-on-V}.
The CO emission in the open spiral arms nicely corresponds to the dust lanes.

These two small spiral arms show a faint extension with a 
progressively smaller
pitch angle, winding up onto a pseudo-ring of radius $\sim$ 9\arcsec=430~pc.
The ring is at the edge of the FoV, and is therefore stronger
in reality than it appears in our map. This ring might correspond to the inner Lindblad resonance of the
nuclear bar (e.g. Buta \& Combes 1996).
Such a ring has also been identified by Comeron \etal (2010) at
9.7\arcsec=465~pc radius from the HST images, and corresponds to weak star forming regions
in ionised gas (Comte \& Duquennoy 1982; Ag\"uero \etal 2004).
 The main star-forming ring is, however, at the end of the bar, 
at about 1.5~kpc (Elmegreen \& Elmegreen 1990).

\begin{figure}[h!]
\centerline{
\includegraphics[angle=-90,width=8cm]{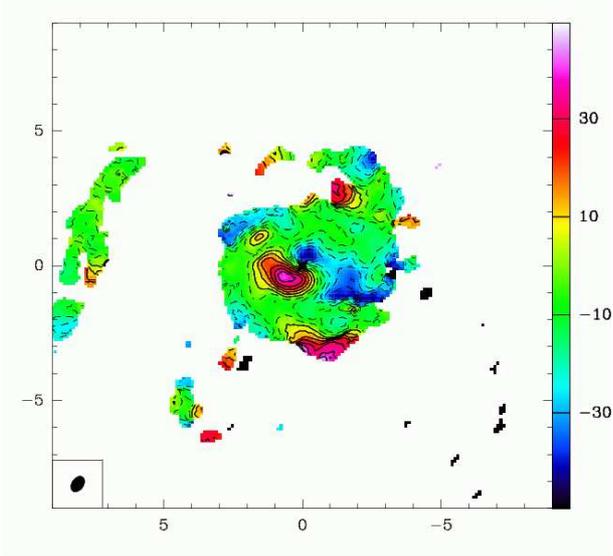}
}
\caption{Velocity residuals after subtraction of a regular rotation model,
based on the CO data, as displayed in Figure \ref{fig:vrot}. The beam size of
0.64\arcsec\ x 0.43\arcsec\ (30 x 20~pc) is indicated at lower left.
The black cross indicates the adopted center.}
\label{fig:resvel}
\end{figure}

\begin{figure}[h!]
\centerline{
\includegraphics[angle=-90,width=8cm]{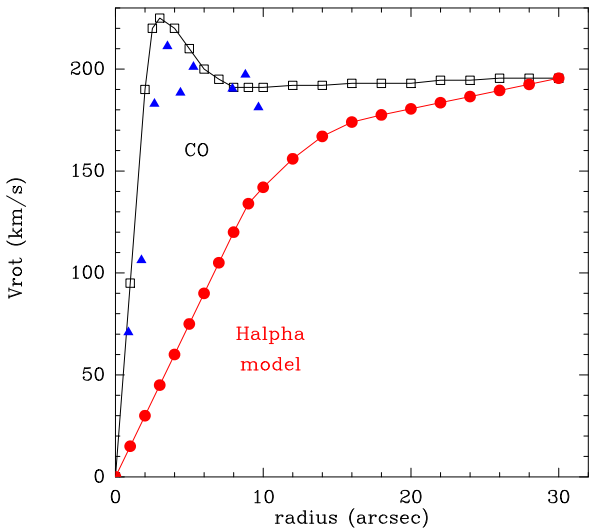}
}
\caption{Rotational velocity model adopted for NGC~1566 (black open rectangles), 
based on the CO data points (blue triangles), and minimizing the residuals of Fig. \ref{fig:resvel}.
The  H$\alpha$ kinematic model from Ag\"uero \etal (2004) is also shown in red circles. 
Those data were obtained with 2\farcs8-wide slits, with insufficient spatial resolution to
resolve the central peak. The H$\alpha$ velocity model rises linearly because of the deficiency of 
HII regions inside a radius of 22\arcsec\ (see text).
}
\label{fig:vrot}
\end{figure}

\subsection{Continuum emission}
\label{cont}

Continuum emission was detected at 0.87mm, using all four sub-bands,
once the lines were subtracted. The bandwidth available then amounts to 6500 MHz,
yielding an rms noise level of 0.055 mJy.
Fig. \ref{fig:cont} displays the CO(3-2) contours superposed on the
continuum map.  The $\sim$ 3mJy peak emission is detected at 50$\sigma$ significance.
 The emission is extended, and follows the 2-arm spiral structure
already detected in the CO(3-2) line. The dust emission has already been 
detected at many wavelengths in this galaxy, with IRAS, ISO and {\it Spitzer}. It peaks
at 102 Jy at 160$\mu$m (Dale \etal 2007), and is still 9.1 Jy at 500$\mu$m (Wiebe \etal 2009).
The emission we see at 870$\mu$m  is thus expected to come essentially from dust.
 From the radio emission at cm wavelengths, the slope of the synchrotron emission
can be derived to be -0.8. Extrapolating from the measured point at 5~GHz, the total
synchrotron emission is expected to be 3mJy. On the high frequency side, the extrapolation of a grey-body, 
with a dust opacity  $\propto\nu^2$, gives a flux at 870 microns of 1 Jy. The dust emission is then
dominating the flux at 870 microns.

 The total continuum emission integrated over our FoV amounts to
12.5 mJy, which is consistent with the extrapolation of  the far-infrared SED 
from previous measurements, if account is taken for the missing flux and
the continuum extent: the total flux for the whole galaxy should be
of the order of 1 Jy. Most of the emission must come from 
outside the ALMA FoV, and it is likely that the interferometer misses 
a significant part of the emission, more so than for the line where the 
velocity splitting reduces the extent of the emission, in each channel.

The central peak in the continuum emission, coinciding with the peak in the
CO(3-2) emission at the center of the nuclear spiral arm,
is adopted as the new center of the galaxy (cf Table \ref{tab:basic}).
The distance between the two centers is $\sim$ 0.5\arcsec\ = 24~pc, and might  
result from the uncertainty of the optical position because of dust patches
at the center.

\begin{figure}[h!]
\centerline{
\includegraphics[angle=0,width=8cm]{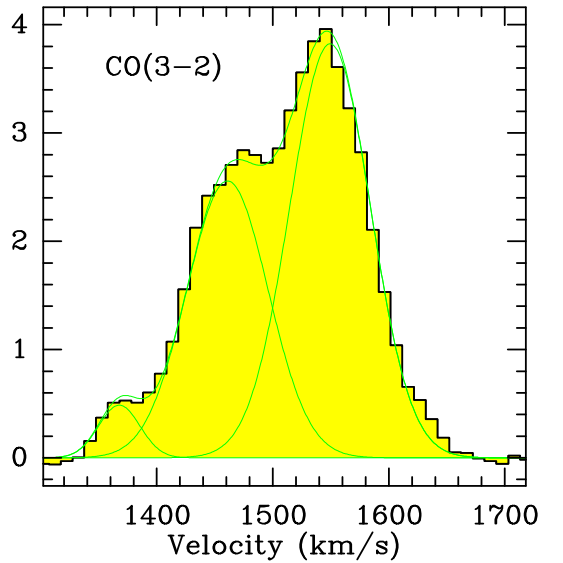}
}
\caption{Total CO(3-2) spectrum, integrated over the observed map,
with a FoV of 18\arcsec. The vertical scale is in Jy. The green line is the result
of the gaussian fit with 3 velocity components (see Table \ref{tab:line}).
The spectrum is primary beam corrected.
}
\label{fig:spectot}
\end{figure}

\subsection{CO kinematics}
\label{COkin}

The top panel of Figure \ref{fig:velo} displays the velocity field of
the molecular gas.  The velocity field is well described by rotation, with the major axis
at a position angle of 44$^\circ$ and an inclination of 
35$^\circ$ as determined by optical studies at larger scales. 
 The CO velocities reveal a peak in the rotation curve at about 2\arcsec, then a slight
decrease to reach a plateau, similar to what is found in optical studies.
  Although the CO rotation curve shown in Fig. \ref{fig:vrot}
seems to be poorly constrained by the scattered CO data points,
 it was more precisely derived by minimising the residuals in the 2D map (see below).

 The H$\alpha$ rotation curve was 
determined by Ag\"uero \etal (2004) from 6 slits of width 2\farcs8 that have 
insufficient spatial resolution to resolve the 2\arcsec\ peak. In addition,
there is a deficiency of
HII regions inside a radius of 22\arcsec = 1~kpc, corresponding to the 
pseudo ring where the grand design spiral arms begin. The Fabry-Perot  H$\alpha$
rotation curves from Comte \& Duquennoy (1982) and Pence \etal (1990) also suffer from
this HII region deficiency, and are consistent with our curve only at larger radii. 
The model H$\alpha$ rotation curve from Ag\"uero \etal (2004) shown in Fig. \ref{fig:vrot}
follows a linear curve inside the ring, as expected for any tracer distribution
with a hole at the center.
The main star forming ring appears at the end of the bar, and might
correspond to the corotation of the bar, coinciding with the inner Lindblad resonance
of the spiral (Elmegreen \& Elmegreen 1990). This would explain the deficiency
of ionised gas inside 1~kpc radius.
Therefore, only the CO gas can probe the disk inside the stellar bar.
The deficiency of star formation in these regions emitting molecular lines
might be due to the dynamical state of the gas, which disfavors star formation,
like in NGC~1530 for example (Reynaud \& Downes 1998).
 
The stellar bar is mainly visible in red and near-infrared images, 
inside the star forming ring.
The molecular gas does not follow the bar, but instead develops a nuclear spiral structure,
which reflects some kinematical perturbations. These are most visible in 
the position-velocity diagram along the minor axis (Fig. \ref{fig:pv-minor}), while the 
rotation is more regular along the major axis (Fig. \ref{fig:pv}).
 These kinematic disturbances arise more from gaseous spiral and turbulent perturbations than from
coherent streaming motions due to a bar. 
 This situation is similar to that in other galaxies in the NUGA sample, which show similar 
wiggles on the minor axis PV diagram but remain regular on the major axis: 
NGC~4826 (Garc\'{\i}a-Burillo \etal 2003), NGC~3147
(Casasola \etal 2008) or NGC~1961 (Combes \etal 2009).
 In NGC~1566, the bar has a position angle of nearly 0$^\circ$, significantly different from the 
major axis position angle of  44$^\circ$. This is a favorable situation for measuring
streaming motions due to the bar. However, the gas follows its own
 spiral structure that is more aligned with the major axis, 
explaining why little perturbation shows up along this symmetry axis.

Another way to reveal  peculiar velocities in 2D over the nuclear region
is to subtract the expected regular velocity field derived from the CO data from
the velocity field.
Fig. \ref{fig:resvel} displays the residuals obtained relative to the adopted
rotation curve model of Fig. \ref{fig:vrot}.  They are very low all over the map, 
and peak at $\sim$ 30 km/s S-SE of the nucleus, corresponding to the 
edge of an arm.  Since at large-scale trailing spiral structures are always dominant,
the observed velocity field indicates that the SE region is the far side, and the NW the near side.
Streaming motions are stronger on the far side, but are present on both sides,
while following the spiral arms. This pattern is not characteristic
of a symmetric outflow of gas. It has the right sign for an outflow, since it is redshifted
on the far side of the minor axis and blueshifted on the near side,
but it has much too small an amplitude. Note also that at larger scales, Ag\"uero \etal (2004) 
find the opposite, i.e. an inflow of gas along the minor axis (see their Fig 3), with
a comparable amplitude. This supports our interpretation in terms of streaming motions.
 The stronger ``outflow" feature coincides in size and position
with the one-sided NLR ionised cone in the [OIII] high resolution image of
Schmitt \& Kinney (1996), but the coincidence is most likely spurious, since
there is no reason for the cone to be related to the spiral arm streaming motion.

\subsection{\hh\, mass}
\label{CO-mass}

Figure \ref{fig:spectot} displays the total CO(3-2) spectrum integrated
over the observed map.  When integrated over the spectral range (FWHM=155km/s), the 
integrated emission is 596$\pm$ 2 Jy km/s.  Towards the central position, Bajaja \etal (1995)
found a CO(2-1) spectrum peaking at T$_A^*$= 120mK with  FWHM=200 km/s,
yielding a total integrated flux of 660 Jy km/s, in a beam of 22\arcsec. Since
their beam is very similar to our FoV, the comparison is meaningful, telling us that 
we are certainly missing some flux, by a factor up to $\sim$ 2.
Indeed, the CO excitation appears high at the center of NGC~1566, since the CO(2-1)/CO(1-0) 
intensity ratio is about 1 in temperature units, averaged over the 43\arcsec\ beam (Bajaja \etal 1995). 
This is expected for thermalized
excitation and a dense molecular medium. In that case the CO(3-2) flux should be higher than
the CO(2-1) by up to a factor of 2.2 (flux varying as $\sim \nu^2$ for gas at temperature larger than 25K
and density larger than 10$^4$ cm$^{-3}$).
 Already the CO(2-1) spectrum of Bajaja \etal (1995) appears broader in velocity,
which indicates some missing flux in our CO(3-2) map.
 The SEST observations point to a total molecular mass of 1.3 10$^9$ M$_\odot$, and 
in the central 43\arcsec\ beam of  3.5 10$^8$ M$_\odot$. In the 22\arcsec\ beam, the SEST CO(2-1)
spectrum, together with the CO(2-1)/CO(1-0) ratio of 1, yields a mass  1.7 $\times$ 10$^8$ M$_\odot$, 
while we find  0.7 $\times$ 10$^8$ M$_\odot$ in our
FoV of 18 \arcsec\, assuming  thermally excited gas, and a Milky-Way like CO-to-H$_2$ conversion factor,
of 2.3 10$^{20}$cm$^{-2}$/(Kkm/s)  (e.g. Solomon \& Vanden Bout 2005).

\subsection{HCO$^+$ and HCN}
\label{HCO-HCN}

\begin{figure}[h!]
\centerline{
\begin{tabular}{c}
\includegraphics[angle=-90,width=8cm]{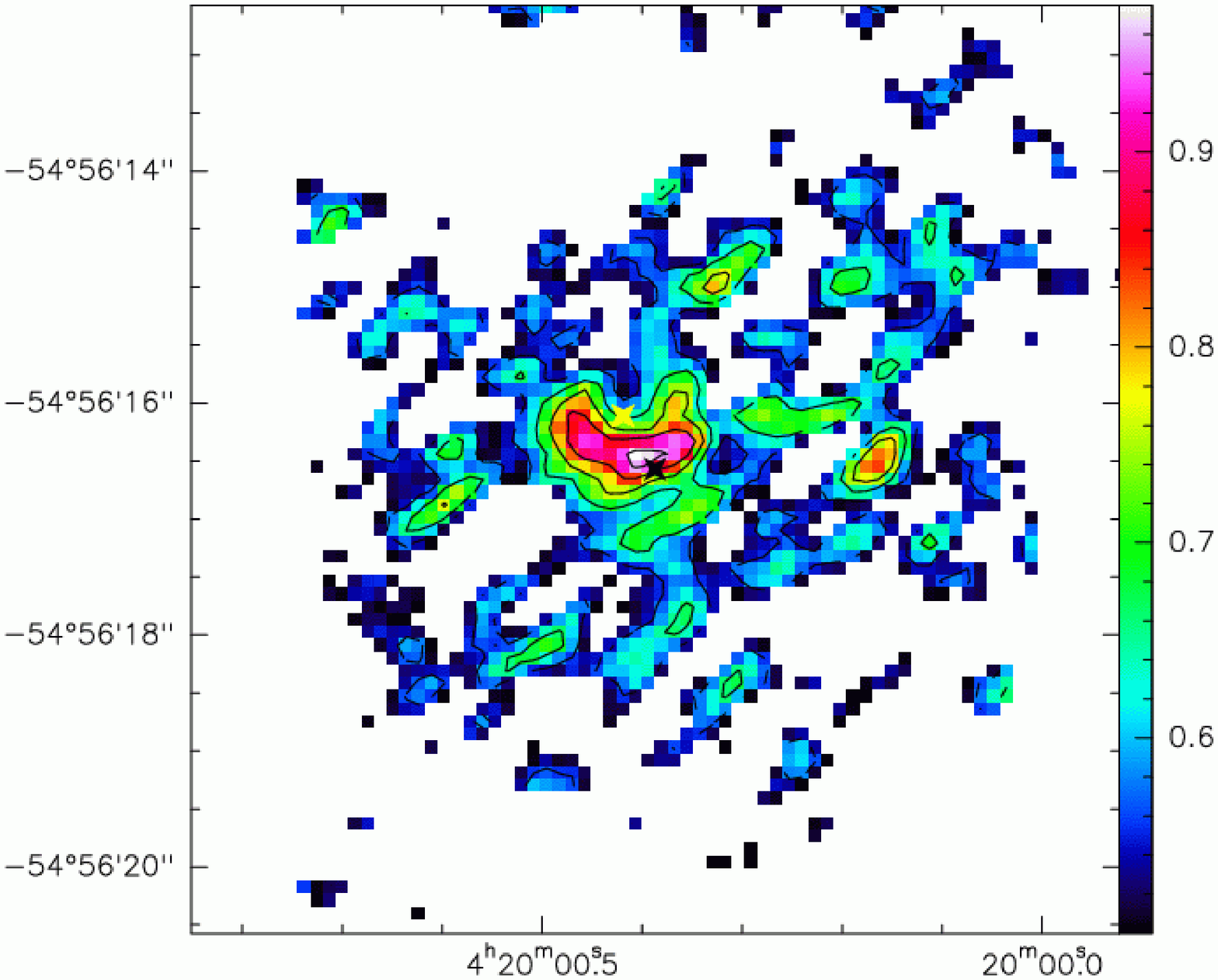}\\
\includegraphics[angle=-90,width=8cm]{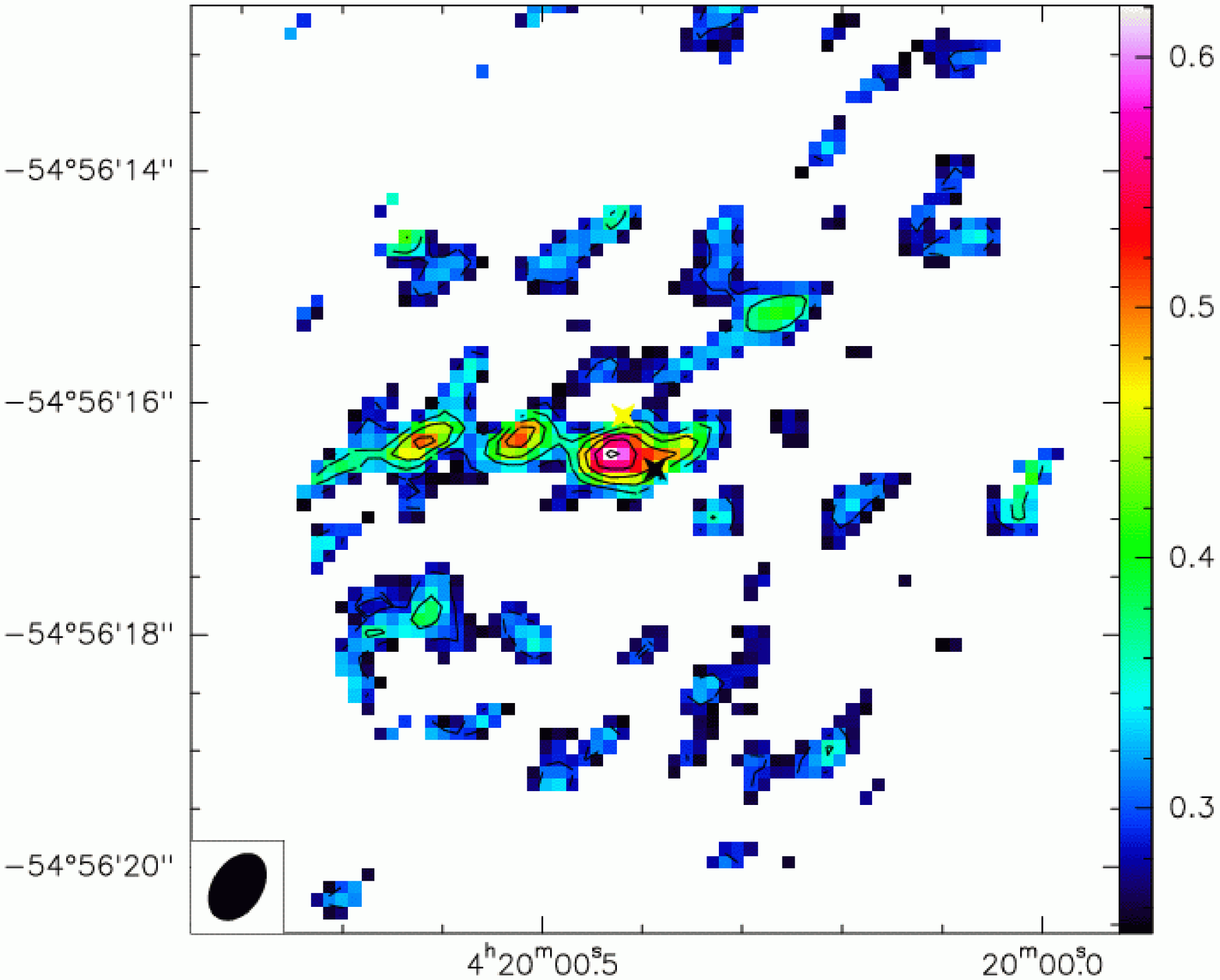}\\
\end{tabular}
}
\caption{{\it Top}:  Map of the HCO$^+$(4-3) line in NGC~1566. The adopted center
is drawn with a black cross, and the phase center by a yellow cross. The color scale is in Jy/beam$\times$MHz
(or 0.87 Jy/beam$\times$km/s).
{\it Bottom:} Same for HCN(4-3). The beam size of 30 x 20~pc is indicated at the lower left.
}
\label{fig:HCO-HCN}
\end{figure}

\begin{figure}[h!]
\centerline{
\includegraphics[angle=0,width=4cm]{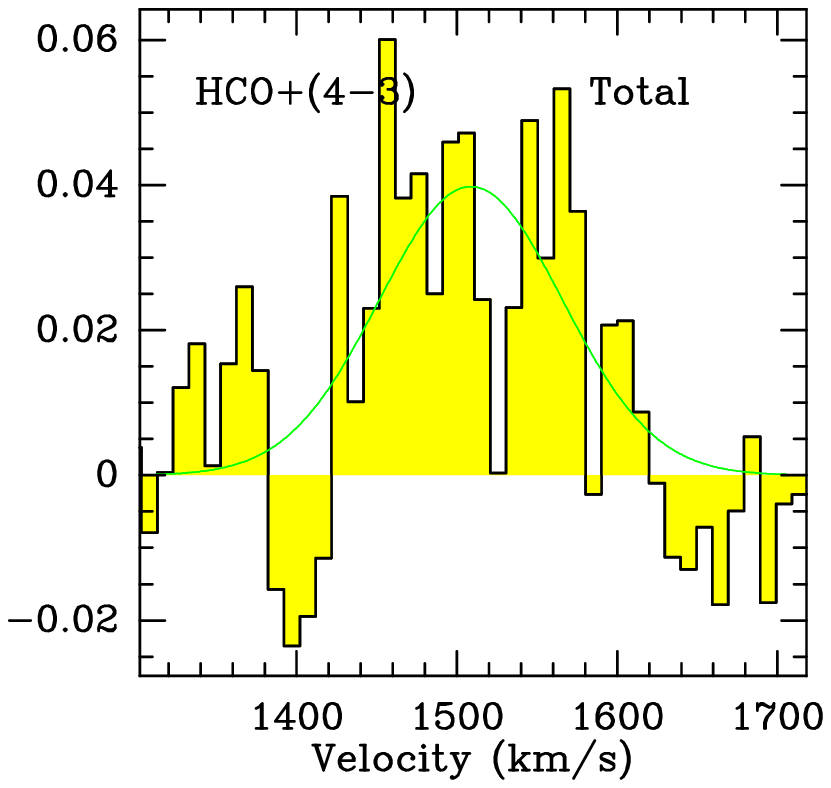}
\includegraphics[angle=0,width=4cm]{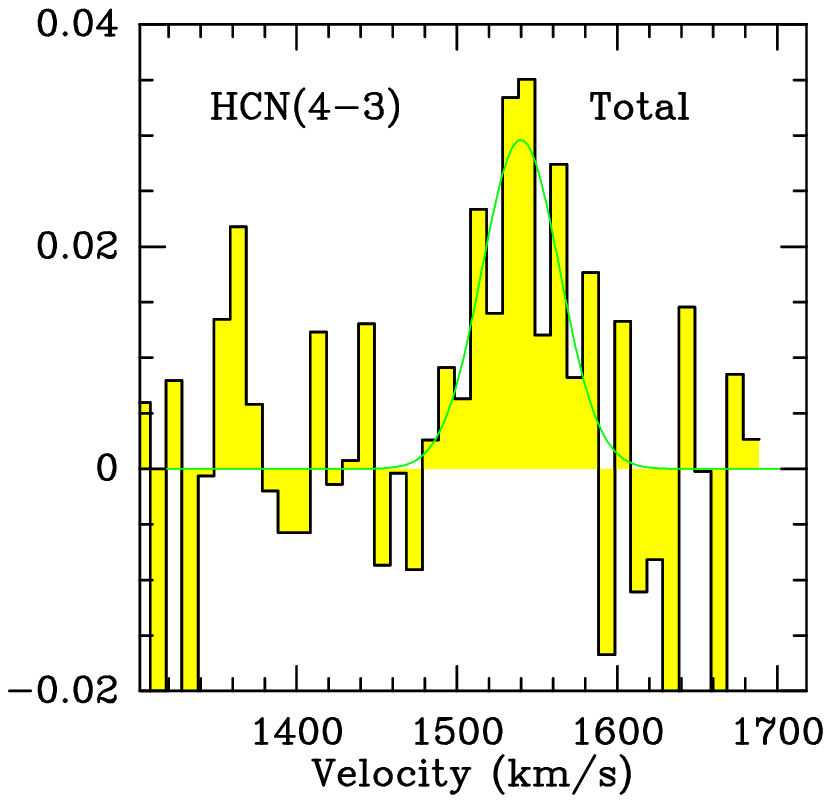}
}
\caption{Total HCO$^+$(4-3){\it (left)} and HCN(4-3) {\it (right)} spectra, 
integrated over the field of view of 18\arcsec, after primary beam correction. 
The vertical scale is in Jy. The green lines are
 the Gaussian fits, whose parameters are displayed in Table \ref{tab:line}.
}
\label{fig:specHCO}
\end{figure}

Along with CO(3-2), we  detect the HCO$^+$(4-3) and HCN(4-3) emission lines, 
mainly in the center of the galaxy. The corresponding maps are displayed in Figure \ref{fig:HCO-HCN},
and the integrated spectra in Figure \ref{fig:specHCO}.
The lines are mainly detected in the nuclear spiral structure, and are too faint to see
the rest of the nuclear disk. The HCO$^+$ line is 3 times stronger than the HCN line, which is 
expected for starburst galaxies such as M82 (e. g. Seaquist \& Frayer 2000).
 The detection of these lines reveals a high proportion of dense gas, since the critical densities
to excite the J=4-3 transition of HCO$^+$ and HCN are 6.5 10$^6$ and 1.6 10$^8$ cm$^{-3}$, respectively. 

The integrated intensities and profile characteristics displayed in Table \ref{tab:line} reveal that 
the HCO$^+$ line width is comparable to the CO(3-2) width, while the HCN spectrum appears
truncated, as if only one side of the galaxy, the redshifted SW part, is strong
enough to be detected. The total intensities of both lines are, however, weak:  HCO$^+$(4-3) is
100  times weaker than  CO(3-2), and  HCN(4-3) is 300 times weaker than CO(3-2).
These ratios are about 3 times less than at the center of M82, and suggest a lower
fraction of dense gas (e.g. Naylor \etal 2010). This is expected since NGC~1566 is not
a starburst galaxy (e.g., Gao \& Solomon 2004).

The fact that the HCO$^+$ emission is 3 times stronger than the HCN emission indicates that the excitation
of the molecules at the center of NGC~1566 is dominated by star formation (PDR or
photodissociation region), and not the AGN (XDR or X-ray dominated region).
Indeed, in AGN-dominated molecular disks, the HCN/HCO$^+$ ratio can reach values much larger than 1
(Kohno \etal 2003; Krips \etal 2008; Garc\'{\i}a-Burillo \etal 2010; Costagliola \etal 2011; Imanishi \& Nakanishi 2013).

\section{Torques and AGN fueling}
\label{torq}

  A small bar is detected in red and near-infrared images of NGC~1566
(see for example Fig. 1 from Ag\"uero \etal 2004, from 2MASS).
Its position angle is PA=0$^\circ$, and its length is 35\arcsec= 1.7~kpc,
as determined by Hackwell \& Schweizer (1983), or Comeron \etal (2010).
  According to Elmegreen \& Elmegreen (1990), this radius coincides with 
kinks in the spiral arms, and the amplitude of the conspicuous 2-arm
grand design spiral begins to increase.  We have used a red image (F814W) 
from HST to derive the stellar potential in the center, since 2MASS images
have insufficient angular resolution. We have not separated the bulge from the disk contribution
since NGC~1566 is a late type (Sbc) galaxy. This means that the bulge was effectively 
assumed to be flattened. Dark matter can be safely
neglected inside the central kpc. The image has been rotated and deprojected according
to PA=44$^\circ$ and i=35$^\circ$, and then Fourier transformed to compute the gravitational
potential and forces. A stellar exponential disk thickness of $\sim$1/12th of the radial scale-length of the galaxy
(h$_{\rm r}$=3.8kpc) has been assumed, giving h$_{\rm z}$=317pc. This is the
average scale ratio for galaxies of this type (e.g., Barteldrees \& Dettmar 1994;
Bizyaev \& Mitronova 2002, 2009).
The potential has been obtained assuming a constant mass-to-light ratio
of M/L = 0.5 \msol/\lsol\ in the I-band
over the considered portion of the image of 2~kpc in size. This value is realistic in view
of what is found statistically for spiral galaxies (Bell \& de Jong 2001).
The pixel size of the map is 0.079\arcsec=3.8~pc.
  The stellar M/L value was fit to reproduce the observed CO rotation curve.
 We have checked that the results are not significantly changed when
the geometrical parameters (i and PA) are varied by $\pm$5$^\circ$, as found 
in the literature. The M/L varies as 1/sin$^2$i accordingly.

\begin{figure}[h!]
\centerline{
\includegraphics[angle=-90,width=7cm]{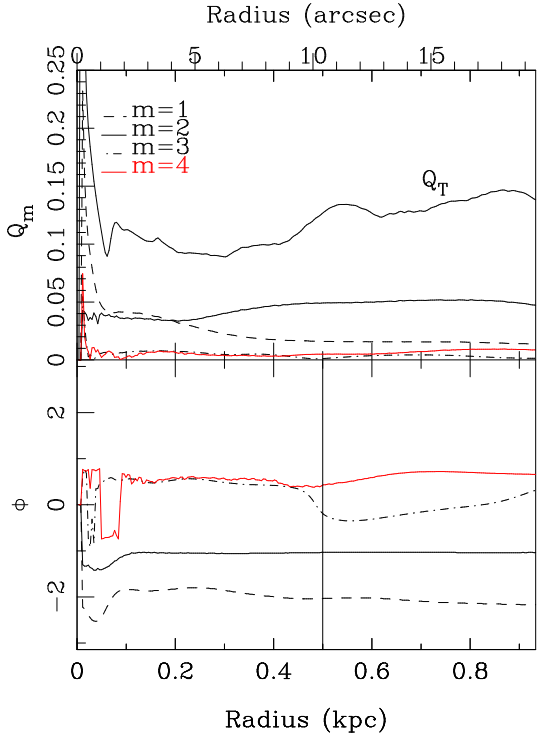}
}
\caption{{\it Top} Strengths (Q$_m$  and total Q$_T$) of  the $m=1$ to $m=4$ Fourier
  components of the stellar potential within the central kpc. The
$m=2$ term is dominant, and has a constant phase, corresponding to the bar.
{\it Bottom} Corresponding phases in radians of the Fourier components, taken from the major axis, 
in the deprojected image. The vertical
line indicates the proposed ILR position for the nuclear bar.}
\label{fig:pot1566}
\end{figure}

The potential $\Phi(R,\theta)$ can be decomposed into its different Fourier components:
  $$
  \Phi(R,\theta) = \Phi_0(R) + \sum_m \Phi_m(R) \cos (m \theta - \phi_m(R))
  $$
  \noindent
  The strength of the $m$-Fourier component, $Q_m(R)$ is defined as
  $Q_m(R)=m \Phi_m / R | F_0(R) |$, i.e. by the ratio between tangential
  and radial forces (e.g. Combes \& Sanders 1981).
  The strength of the total non-axisymmetric perturbation $Q_T(R)$ is defined 
similarly with the maximum amplitude of the tangential force $F_T^{max}(R)$.
  Their radial distributions and the radial phase variations are displayed in Fig.~\ref{fig:pot1566}.

\begin{figure}[h!]
\centerline{
\includegraphics[angle=0,width=8cm]{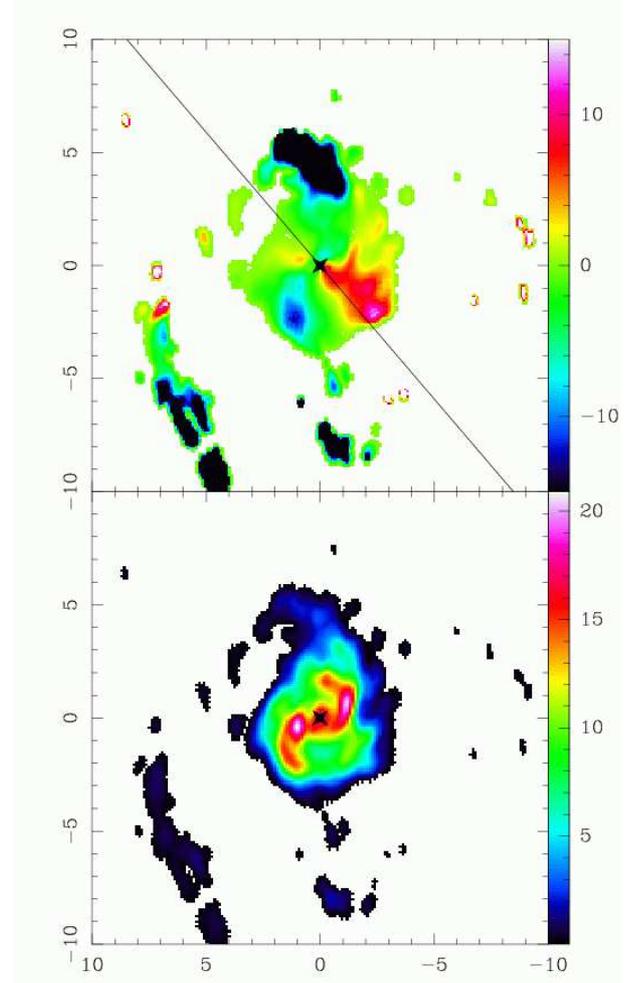}
}
\caption{{\it Top:} Map of the gravitational torque,  (t(x,y)~$\times$~$\Sigma$(x,y), as defined in text) 
in the center of NGC~1566.
  The torques change sign as expected in a four-quadrant pattern
(or butterfly diagram). The orientation of the quadrants follows
  the nuclear bar's orientation. In this deprojected picture,
  the major axis of the galaxy is oriented parallel to the horizontal axis.
  The inclined line reproduces the mean orientation of the bar
  (PA = 40$^\circ$ on the deprojected image). 
{\it Bottom:} The deprojected image of the CO(3-2) emission, at the same scale,
and with the same orientation, for comparison. The axes are labelled in arcsecond relative to the center.
The color scales are linear, in arbitrary units.}
\label{fig:torq1}
\end{figure}

 The derivatives of the potential yield the forces per unit mass ($F_x$ and $F_y$) at
  each pixel, and the torques per unit mass $t(x,y)$ are then computed by:
  $$
  t(x,y) = x~F_y -y~F_x
  $$
  The sign of the torque is determined relative to the sense of rotation in the plane of the galaxy.
  The product of the torque and the gas density $\Sigma$  at each pixel  allows one
  then to derive the net effect on the gas, at each radius.  This quantity $t(x,y)\times
  \Sigma(x,y)$, is shown in Fig.~\ref{fig:torq1}.

\begin{figure}[h!]
\centerline{
\includegraphics[angle=-90,width=7cm]{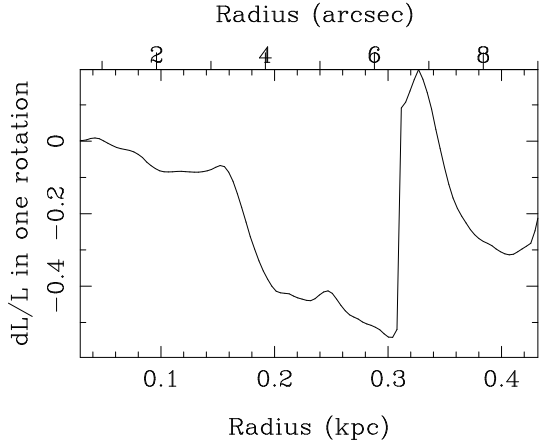}
}
\caption{The radial distribution of the torque, quantified by the fraction of the angular momentum transferred
  from the gas in one rotation--$dL/L$, estimated from the CO(3-2) deprojected map. The forces at 
each point of the map were computed from the I-band HST image (see text).
The torque is negative inside a 0.3 kpc radius and down to the spatial resolution of the observations.
Between 6 and 9\arcsec\ from the center, there is not much CO emission and the torque estimation is noisy.}
\label{fig:gastor}
\end{figure}

  The torque weighted by the gas density $\Sigma(x,y)$ is then averaged over azimuth, i.e.
  $$
  t(R) = \frac{\int_\theta \Sigma(x,y)\times(x~F_y -y~F_x)}{\int_\theta \Sigma(x,y)}
  $$
\noindent
  The quantity $t(R)$ represents the time derivative of the specific angular momentum $L$ of the gas averaged
  azimuthally (e.g. Garc\'{\i}a-Burillo \etal 2005).
 Normalising at each radius by the angular momentum and rotation period
allows us to estimate the efficiency of the gas flow, as shown in   Fig.~\ref{fig:gastor}.

Although the nuclear bar strength is moderate,  Fig.~\ref{fig:gastor} shows that its fueling 
efficiency is high. Between 200 and 300~pc, the gas loses half of its angular momentum in one rotation, 
which is $\sim$  7 Myr. Inside 100~pc, the relative loss rate is lower, but other dynamical phenomena
can then contribute, such as dynamical friction of molecular clouds against the 
distribution of stars. The rotation period becomes
smaller than 3Myr, and the dynamical friction time-scale is decreasing rapidly, especially for
massive giant molecular clouds. Since dense clouds exist at the center of NGC~1566, as 
supported by the HCN and HCO$^+$ detections, this might be the mechanism for driving the
gas down to 10~pc scales, in times comparable to a few rotation periods.
 The dynamical friction time-scale t$_{df}$ at a radius R can be estimated in a galaxy like the Milky Way as
t$_{df}$ (Myr) = (R/10pc)$^2$ (10$^6$ \msol/M$_{cloud}$) for a molecular cloud of mass
M$_{cloud}$ (e.g., Combes 2002).

As shown in Fig. \ref{fig:torq1}, the nuclear spiral structure inside the nuclear ILR ring 
of the bar is of a trailing nature and is located inside the negative torque quadrants. This  
might appear suprising, since in many cases the spiral structure is predicted to be leading
in this region, and the torque positive, maintaining the gas in the ILR ring (e.g., Buta \& Combes 1996). 
However, in  the presence of a sufficiently massive black hole, this behaviour can be reversed: the 
spiral becomes trailing, and the torque negative.
Indeed, the indicator of the precession rate of elliptical orbits in the frame of the 
epicyclic approximation, $\Omega-\kappa/2$, is significantly modified by a central
massive body. Instead of decreasing regularly towards zero at the center, the $\Omega-\kappa/2$  curve
increases steeply as r$^{-3/2}$. The gas undergoes collisions, loses energy and spirals
progressively towards the center. When the precession rate decreases, the series of elliptical 
orbits precess more and more slowly, and then lag at smaller radii,  forming a leading structure.
 When the precession rate increases, they form a trailing structure.

Another question is whether the orbits change orientation at the ILR, as expected if the x$_2$
orbits develop between two inner resonances. This would destroy the nuclear bar and be observable
on the infrared image. The lower panel of Fig.~\ref{fig:pot1566} shows the phases of the various Fourier components.
The bar ($m=2$) has a constant phase over its length, and their orientation change does not appear
to be significant.

\begin{figure}[h!]
\centerline{
\includegraphics[angle=-90,width=7cm]{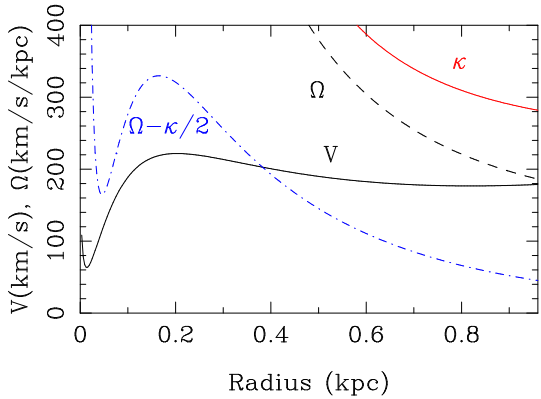}
}
\caption{Schematic representation of the rotation curve (black)
epicyclic frequency $\kappa$ (red), and corresponding $\Omega-\kappa/2$ curve (blue)
within the central kpc of NGC~1566. The contribution of a super-massive black
hole in the nucleus with M$_{\rm BH}$ = 8.3 10$^6$ \msol\ has been included.}
\label{fig:vbh}
\end{figure}

NGC~1566 is therefore an example of  a trailing spiral inside
the nuclear ring of a bar i.e., the case described in Fig 79b of Buta \& Combes (1996).
This means that the mass of the black hole should be sufficiently high to have 
an influence on the gas dynamics on a 100~pc scale. Woo \& Urry (2002) have estimated
a BH mass of M$_{\rm BH}$ = 8.3 10$^6$ \msol\ from the M-$\sigma$ relation using a measured
central velocity dispersion (Table \ref{tab:basic}). With this plausible black hole mass,
we have estimated the BH influence. Although the modification of the rotation curve is 
very small, the influence on the precession rate  $\Omega-\kappa/2$  is more significant.
Figure \ref{fig:vbh} illustrates more clearly how this is possible.

In this figure, we selected an analytical form for the rotation curve that 
schematically reproduces the observed one. It is not possible to make a better fit,
given that only the CO emission is able to trace the kinematics in this region.
 If we locate corotation at the end of the bar, i.e. at a radius of 1.7~kpc, then 
the pattern speed of the bar is $\sim$ 120 km/s/kpc. The ILR is then approximately
at a radius of 500~pc, near the location where we found the resonant ring (see  Sec. \ref{morpho}
and also Comeron \etal 2010). This is a good fit, given that the computation of resonances via
the $\Omega-\kappa/2$ curve and the epicyclic approximation is not accurate.
 The precession rate of the infalling orbits is then going to increase towards the center,
accounting for the trailing spiral structure observed.

Fig. \ref{fig:vbh} shows that the influence of the black hole on the rotation 
curve is almost negligible; it is significant only inside 13~pc. However, it
is more important in influencing the precession rate, which includes the second derivative of $\Omega$. 
The influence of the BH is then felt up to 50~pc.
Since the 100~pc-scale arms are defined by elongated orbits extending below
50~pc, it is possible that the black hole indeed has an influence on the 2-arm spiral structure
inside the ILR ring.

\subsection{Comparison with other nuclear patterns}
\label{compar}

The observation of a trailing 2-arm spiral structure inside the ILR
of a nuclear bar is a rare configuration. 
In molecular gas morphologies, this case had not previously been encountered.
 Although previous NUGA studies have not had the advantage of as small a physical resolution as the current ALMA data, it is useful to make some
statistical considerations. 
Looking at previous NUGA galaxies where detailed torques have been measured,
it is possible to distinguish at least three generic configurations
(see also Garc\'{\i}a-Burillo \& Combes 2012):

\begin{enumerate}

\item There is no ILR, and the trailing structure goes right to the 
nucleus from corotation (CR).
A trailing spiral structure is then expected throughout, with negative torques 
(as in NGC~3147, 3627).
In some cases, like NGC~6574, there is an ILR, but it cannot be resolved 
with the CO beam; negative torques are measured from CR to ILR.
So this case is also similar to the no-ILR case.

\item  There is an ILR, and a ring or pseudo-ring at the ILR,
which is resolved by the molecular observations. The molecular gas
reveals a hole inside the ring, or is highly deficient inside that radius. 
The torque is positive inside the ILR (negative outside), so the gas is 
piling up in the ring.
When there is some CO emission inside the ring, its distribution
is in general patchy, with no clear arms; it could be a
mixture of leading and trailing features, but there must be some 
leading pattern to account for positive torques
(this is the case for NGC~4321, 4569, 5248, 5953, 6951 and 7217).
NGC~1433 is also in this configuration, although the nuclear spiral structure is rather
flocculent (Combes \etal 2013). The spiral is trailing between CR and ILR of the nuclear bar,
and within the ILR at 200~pc in radius, the spiral structure is unclear, but the torque
is positive. 

\item  There is an independent embedded structure,
a nuclear bar with a different pattern speed from that of the primary bar, and 
a negative torque can be measured inside the CR of the nuclear bar, but without a
coherent orientation (case of NGC~2782).

\end{enumerate}

\noindent There are also other special cases, e.g. involving development of 
 a dominant $m=1$ pattern in the center,
like NGC~1961, 4826, or 5850.

Another way to detect spiral structure in the nuclear gas is to amplify
the dust extinction features in the HST images through unsharp masking. 
Martini \etal (2003a,b) have studied
the presence of circumnuclear dust in 123 galaxies, of which 14 have nuclear rings.
When circumnuclear dust features are present, they have classified them
as grand design 2-arm structures (GD), multi-arm or floculent structures
(tightly or loosely wound) and chaotic structures (spiral or not).
Only 19 galaxies have GD nuclear morphologies, and among the nuclear
ring galaxies, only 3 are classified as GD (NGC~1300, 3081 and 6890).
Only galaxies with large scale bars exhibit  GD nuclear spiral structure.
Most of these spiral structures continue the dust lanes along the bar,
without any ILR.
In their sample of 75 nearby galaxies, Peeples \& Martini (2006) find 7 objects
with GD nuclear spirals. Only two have both a large-scale
grand design and a GD, NGC~3081 and 5643. In these galaxies, however,
the sense of winding of the dust spirals in the very center is unclear,
and there could be a mixture of leading and trailing features. A clear case of 
a trailing spiral inside a nuclear ring is still missing.
We have also  examined the
compilation of nuclear rings of Comeron \etal (2010).
Most rings are either devoid of structure, or display confused or chaotic dust structures.
Although the nuclear ring is clearly seen in the NGC~1566 image, the interior structure
is difficult to distinguish in dust maps, making the molecular gas emission a better tracer.

\section{Summary}
\label{disc}

We have presented ALMA cycle 0 results for a Seyfert 1 galaxy from our extended NUGA sample, NGC~1566.
The observations in CO(3-2) allow us to reach a superb spatial resolution of 25 pc, even
in this early cycle.

The center of NGC~1566 has a dense nuclear disk of molecular gas, that is deficient in both
atomic gas and HII regions. The central disk is typically $\sim$ 3.5\arcsec=170~pc in radius, and
reveals an open 2-arm nuclear spiral structure, which tends to wind up into a faint
ring at 430~pc radius. The spiral structure coincides very well with obscured
dust lanes in HST images, and also with dust continuum emission at 870$\mu$m.
The nuclear spiral arms are dense enough to be detected in the 
HCO$^+$(4-3) and HCN(4-3) lines.

The kinematics of the CO emission show a rather regular rotational velocity field,
with only slight perturbations from the 2-arm nuclear spiral. There are in particular
redshifted streaming motions on the SE part of the minor axis, and blueshifted ones
on the NW side, along the border of the spiral arms.
  However, the amplitude of these perturbations is small, and it is remarkable that 
no symmetric molecular outflow, indicating a possible AGN feedback,
is observed.

The faint CO ring at r=430~pc, which is more strongly noticeable in the V-I colour
HST image (Comeron \etal 2010), is plausibly stronger in reality given the reduced sensitivity
at the edge of our FoV. This ring corresponds to the ILR of the bar, if corotation is 
assumed to coincide with the bar end.
  The trailing spiral observed in CO emission is then well inside the ILR ring
of the bar. We have computed the gravitational potential from the stars
within the central kpc, from the I-band HST image. Weighting the torques
on each pixel by the gas surface density observed in the CO(3-2) line has 
allowed us to estimate the sense of the angular momentum exchange
and its efficiency.  The gravity torques are negative from 50 to 300pc. 
Between 200~pc and 300~pc, the gas loses half
of its angular momentum in a rotation period. Gravity torques are
then very efficient in fueling the AGN, and bring molecular clouds 
so close to the center that dynamical friction could drive them
to the nucleus in a few Myr.

This work provides the
first example of a clear trailing spiral observed in molecular gas inside
the ILR ring of a bar, indicating that the super-massive black hole
is influencing the gas dynamics significantly enough to reverse
gravity torques. Instead of maintaining the ILR ring density,
the torques are then driving
the gas towards the nucleus, a first step towards possibly fueling the AGN.

\begin{acknowledgements}
The ALMA staff in Chile and ARC staff at IRAM are gratefully acknowledged for their
help in the data reduction. We want to thank in particular Gaelle Dumas and Philippe
Salom\'e for their useful advice.
This paper makes use of the following ALMA data: ADS/JAO.ALMA\#2011.0.00208.S.
ALMA is a partnership of ESO (representing its member states), NSF (USA) and NINS (Japan),
together with NRC (Canada) and NSC and ASIAA (Taiwan), in cooperation with the Republic of
Chile. The Joint ALMA Observatory is operated by ESO, AUI/NRAO and NAOJ.
The National Radio Astronomy Observatory is a facility of the National Science Foundation
operated under cooperative agreement by Associated Universities, Inc.
We used observations made with the NASA/ESA {\it Hubble} Space Telescope, and obtained
from the {\it Hubble} Legacy Archive, which is a collaboration between the Space Telescope
Science Institute (STScI/NASA), the Space Telescope European Coordinating Facility
(ST-ECF/ESA), and the Canadian Astronomy Data Centre (CADC/NRC/CSA).
F.C. acknowledges the European Research Council
for the Advanced Grant Program Num 267399-Momentum.
I.M. acknowledges financial support from the Spanish grant AYA2010-15169
and from the Junta de Andalucia through TIC-114 and the Excellence Project
P08-TIC-03531.
We made use of the NASA/IPAC Extragalactic Database (NED),
and of the HyperLeda database.
\end{acknowledgements}


\begin{thebibliography}{}
\bibitem{}Abramowicz, M. A., Xu, C., Lasota, J. P. 1986 IAU, S119, 371
\bibitem{}Ag\"uero E.L., Diaz R.J., Bajaja E., 2004, A\&A 414, 453 
\bibitem{}Alatalo K., Blitz L., Young L.M.  et al,  2011, ApJ 735, 88
\bibitem{}Alloin, D., Pelat, D., Phillips, M. M., Whittle M., 1985 ApJ 288, 205 
\bibitem{}Alloin, D., Pelat, D., Phillips, M. M., Fosbury, R. A. E., Freeman, K., 1986 ApJ 308, 23 
\bibitem{}Bajaja E., Wielebinsli R., Reuter H.P. \etal, 1995, A\&AS 114, 147 
\bibitem{}Baribaud, T., Alloin, D., Glass, I., Pelat, D., 1992 A\&A 256, 375
\bibitem{}Barteldrees A., Dettmar R-J., 1994 A\&AS 103, 475
\bibitem{}Bell E.F., de Jong R.S., 2001, ApJ 550, 212  
\bibitem{}Buta R., Combes F., 1996, Fal of Cosmics Physics 17, 95
\bibitem{}Bizyaev, D., Mitronova, S. 2002, A\&A, 389, 795
\bibitem{}Bizyaev, D., Mitronova, S. 2009, ApJ 702, 1567
\bibitem{}Casasola, V., Combes, F., García-Burillo, S. \etal, 2008, A\&A 490, 61 
\bibitem{}Comeron, S., Knapen, J. H., Beckman, J. E. \etal, 2010, MNRAS 402, 2462 
\bibitem{}Combes F., Sanders R.~H., 1981, A\&A 96, 164
\bibitem{}Combes F. 2002, in 7th cosmology colloquium, High Energy Astrophysics, for and from Space, (arXiv:astro-ph/0208113) 
\bibitem{}Combes F.,  Baker, A. J., Schinnerer, E. \etal, 2009, A\&A 503, 73  
\bibitem{}Combes F., Garc\'{\i}a-Burillo S., Casasola V. \etal 2013, A\&A 558, A124 
\bibitem{}Comte, G., Duquennoy, A., 1982, A\&A 114, 7 
\bibitem{}Costagliola F., Aalto S., Rodriguez M.I \etal,  2011, A\&A 528, A30 
\bibitem{}Croton D.J., Springel V., White S.D.M. \etal,  2006 MNRAS 365, 11
\bibitem{}Dale, D.A., Gil de Paz, A., Gordon, K.D. \etal,  2007, ApJ 655, 863 
\bibitem{}Di Matteo, T., Colberg, J., Springel, V., Hernquist, L., Sijacki, D., 2008, ApJ 676, 33
\bibitem{}Elmegreen B.G., Elmegreen D.M., 1990, ApJ 355, 52 
\bibitem{}Feruglio, C., Maiolino, R., Piconcelli, E., et al 2010 A\&A 518, L155 
\bibitem{}Gao Y., Solomon P.M., 2004, ApJS 152, 63
\bibitem{}Garc\`{i}a-Burillo, S., Combes, F., Hunt, L. K. \etal, 2003, A\&A 407, 485 
\bibitem{}Garc\`{i}a-Burillo, S., Combes F., Schinnerer E., Boone F., Hunt L.K., 2005 A\&A 441, 1011
\bibitem{}Garc\`{i}a-Burillo, S., Usero A., Fuente A. \etal, 2010, A\&A 519, A2 
\bibitem{}Garc\`{i}a-Burillo, S., Combes F. 2012, JPhCS Vol 372, 012050
\bibitem{}Glass I.S. , 2004 MNRAS 350, 1049 
\bibitem{}Guilloteau, S., Lucas, R.\ 2000, Imaging at Radio through Submillimeter Wavelengths, 217, 299 10.
\bibitem{}G\"ultekin K., Richstone, D. O., Gebhardt, K. \etal, 2009, ApJ 698, 198
\bibitem{}Hackwell, J. A., Schweizer, F., 1983 ApJ 265, 643	
\bibitem{}Harnett J.I., 1984 MNRAS 210, 13
\bibitem{}Harnett J.I., 1987 MNRAS 227, 887
\bibitem{}Ho, L. C., Li, Z-Y, Barth, A. J. \etal, 2011, ApJS 197, 21
\bibitem{}Hopkins, P.F., Quataert, E. ,  2010, MNRAS 407, 1529
\bibitem{}Hopkins, P., Hernquist, L., Cox, T.J. \etal, 2006, ApJS 163, 1 
\bibitem{}Imanishi M., Nakanishi K., 2013, AJ 146, 47 \& 91 
\bibitem{}Jogee S. 2006 Lecture Notes in Physics 693,  143
\bibitem{}Korchagin, V., Kikuchi, N., Miyama, S. M. \etal, 2000, ApJ 541, 565
\bibitem{}Kilborn V.A., Koribalski B.S., Forbes D. \etal,  2005, MNRAS 356, 77 
\bibitem{}Kohno K, Ishizuki S., Matsushita S. \etal, 2003, PASJ 55, L1 
\bibitem{}Krips M., Neri R., Garc\'{\i}a-Burillo S. \etal, 2008, ApJ 677, 262 
\bibitem{}Krips M., Martín, S., Eckart, A. \etal, 2011,  ApJ 736, 37 
\bibitem{}Levenson, N. A., Radomski, J. T., Packham, C.  \etal,  2009 ApJ 703, 390 
\bibitem{}Malkan M., Gorjian V., Tam R., 1998 ApJS 117, 25 
\bibitem{}Martini, P., Regan, M.W., Mulchaey, J.S., Pogge, R.W. 2003a, ApJS 146, 353
\bibitem{}Martini, P., Regan, M.W., Mulchaey, J.S., Pogge, R.W. 2003b, ApJ  589, 774
\bibitem{}McMullin, J.~P.,  Waters, B., Schiebel, D., Young, W., Golap, K.\ 2007, 
Astronomical Data Analysis Software and Systems XVI, 376, 127 
\bibitem{}Naylor B.J., Bradford C.M., Aguirre J.E. \etal, 2010, ApJ 722, 668 
\bibitem{}Pence W.D., Taylor K., Atherton P., 1990, ApJ 357, 415 
\bibitem{}Peeples M.S., Martini P., 2006 ApJ 652, 1097
\bibitem{}Reif, K., Mebold, U., Goss, W. M. \etal, 1982, A\&AS 50, 451 
\bibitem{}Renaud, F., Bournaud, F., Emsellem, E. \etal, 2013, MNRAS 436, 1836
\bibitem{}Reunanen, J., Kotilainen, J. K., Prieto, M. A. 2002, MNRAS 331, 154
\bibitem{}Reunanen, J., Prieto, M. A., Siebenmorgen, R. 2010, MNRAS 402, 879
\bibitem{}Reynaud D., Downes D., 1998 A\&A 337, 671 
\bibitem{}Riffel R.A., Storchi-Bergmann T., 2011, MNRAS  411, 469  
\bibitem{}Rupke, D. S., Veilleux, S., Sanders, D. B.,  2005 ApJ 632, 751  
\bibitem{}Schmitt H.,  Kinney A., 1996 ApJ 463, 498 
\bibitem{}Seaquist E.R., Frayer D.T., 2000, ApJ 540, 765 
\bibitem{}Shlosman, I., Frank, J., Begelman, M. C., 1989, Nature  338, 45
\bibitem{}Sijacki, D., Springel, V., Di Matteo, T., Hernquist, L., 2007, MNRAS 380, 877
\bibitem{}Solomon P.M., Vanden Bout P.A. 2005, ARAA, 43, 677 
\bibitem{}Sturm, E., Gonzalez-Alfonso, E., Veilleux, S. \etal, 2011, ApJ 733, L16 
\bibitem{}Walsh, W., 2004, preprint 
\bibitem{}Wiebe, D. V., Ade, P.A.R., Bock, J. J. \etal, 2009,  ApJ 707, 1809 
\bibitem{}Woo J-H., Urry C.M., 2002, ApJ 579, 530
\end{thebibliography}
\end{document}